\documentstyle[prb,aps,tighten,preprint,eqsecnum]{revtex}

\begin{document}
\title{Distribution of complex eigenvalues for\\
symplectic ensembles of non-Hermitian matrices.}
\author{A.~V.~Kolesnikov$^{1}$and K.~B.~Efetov$^{1,2}$}
\address{$^{1}$Fakult\"{a}t f\"{u}r Physik und Astronomie,\\
Ruhr-Universit\"{a}t Bochum,\\
Universit\"{a}tsstr. 150, Bochum, Germany\\
$^{2}$L.D. Landau Institute for Theoretical\\
Physics, Moscow, Russia}
\date{\today{}}
\maketitle

\begin{abstract}
Symplectic ensemble of disordered non-Hermitian Hamiltonians is studied.
Starting from a model with an imaginary magnetic field, we derive a proper
supermatrix $\sigma $-model. The zero-dimensional version of this model
corresponds to a symplectic ensemble of weakly non-Hermitian matrices. We
derive analytically an explicit expression for the density of complex
eigenvalues. This function proves to differ qualitatively from those known
for the unitary and orthogonal ensembles. In contrast to these cases, a {\it %
depletion} of eigenvalues near the real axis occurs. The result about the
depletion is in agreement with a previous numerical study performed for QCD
models.
\end{abstract}


\section{Introduction}

Properties of non-Hermitian random operators or matrices have attracted
recently a considerable attention. Non-Hermitian random Hamiltonians can
appear as a result of mapping of a model for flux lines in a $\left(
d+1\right)$-dimensional superconductor with line defects in a tilted
magnetic field on a $d$-dimensional model for bosons in a random potential%
\cite{hatano}. Non-Hermitian operators enter Fokker-Planck equations that
describe diffusion and advection of classical particles in a spatially
random but time-independent velocity field \cite
{fisher,aron,kravtsov,bouchaud,chalker} and also determine equations used
for study of problems of turbulence \cite{burgers,fogedby,gurarie}.

Ensembles of random complex non-Hermitian and real asymmetric matrices find
their application for a description of dissipative quantum maps\cite{haake}
in neural network dynamics\cite{sompol}. Recently it was suggested that they
could be relevant for QCD where they correspond to a random Dirac operator
with a non-zero chemical potential\cite{stephanov}. Starting from the first
works\cite{ginibre,girko}, properties of the ensembles of the non-Hermitian
matrices were intensively studied in a considerable number of publications 
\cite{grobe,sommers,janik,zee,fyodorov,fyodorov1}.

Unusual properties of the ensembles of the non-Hermitian operators or
matrices are related to the fact that eigenvalues of the operators and
matrices can be complex. Completely different methods have been used for
study of distributions of the eigenvalues on the complex plane. For example,
the authors of Refs.\cite{chalker,zee} applied Green functions methods while
in Refs.\cite{fyodorov1}, a method of orthogonal polynomials was used. In Ref.%
\cite{fyodorov}, a new regime of a ``weak non-Hermiticity'' was found and the
authors have calculated a joint probability of complex eigenvalues for
complex weakly non-Hermitian matrices. For calculations they used the
supersymmetry technique \cite{book} and derived a zero-dimensional
non-linear $\sigma $-model. An important information about the distribution
function of complex eigenvalues of $N\times N$ matrices for the orthogonal,
unitary and symplectic chiral random matrix ensembles has been obtained
recently numerically\cite{verbaar}.

Although the model with the non-Hermitian Hamiltonian of Ref.\cite{hatano}
differ from those with random matrices, they turn out to be closely related
to each other. In Ref.\cite{efetov}, the model with the non-Hermitian
Hamiltonian $H$ was studied using the supersymmetry method. This Hamiltonian
can be written in the form  \begin{equation} H=\frac{(\hat{{\bf p}}+i
{\bf h})^2}{2m}+U({\bf r)} \; , \label{a1} \end{equation}
where $\hat{{\bf p}}=-i\nabla $, $m$ is the mass of particles, and $U\left( 
{\bf r}\right) $ is a random potential. The vector ${\bf h}$ is proportional
to the component of the magnetic field perpendicular to the direction of the
line defects in the initial problem of the vortices in superconductors. The
Hamiltonian (\ref{a1}) describes a particle moving in an imaginary
vector-potential $i{\bf h}$ and a real random potential $U\left( {\bf r}%
\right) $. The distribution of complex eigenvalues on the complex plane can
be extracted from the distribution function $P(\epsilon ,y)$ defined as
follows \begin{equation}
P(\epsilon ,y)=\frac 1V\left\langle \sum_k\delta (\epsilon -\epsilon
_k^{\prime })\delta (y-\epsilon _k^{\prime \prime })\right\rangle \; , 
\label{a2} \end{equation}
where $\epsilon _k^{\prime }$ and $\epsilon _k^{\prime \prime }$ are the
real and imaginary parts of the eigenenergies, respectively and $V$ is the
volume of the system. The angle brackets stand for an averaging over the
random potential and the sum should be taken over all states.

The problem of calculation of the function $P(\epsilon ,y)$ was mapped in
Ref.\cite{efetov} onto a new supermatrix non-linear $\sigma $-model. This
model differs from the conventional ones written previously\cite{book} by
the presence of new ``effective fields''. The symmetry of the matrix $Q$
entering the new $\sigma $-model is the same as that of obtained in Ref.\cite
{book} for the orthogonal ensemble. This is not accidental because the
Hamiltonian, Eq. (\ref{a1}), is real and, hence, time reversal invariant.

To violate the time reversal invariance one can add to the Hamiltonian (\ref
{a1}) a real magnetic field or/and magnetic impurities. This leads to
additional terms in the $\sigma $-model lowering the symmetry of the model.
As a result, one gets\cite{efetov} the $\sigma $-model with the
supermatrices $Q$ corresponding to the unitary ensemble. Although the real
magnetic interactions in the Hamiltonian with the imaginary vector potential
do not correspond to any physical interactions in the initial problem of the
vortices, consideration of the $\sigma $-model for the unitary ensemble was
interesting from the formal point of view because it allowed to establish
important relations with the random matrix models.

The $\sigma $-models corresponding to the Hamiltonian of Eq. (\ref{a1}) and
its extensions can be written in an arbitrary dimension. Remarkably, the
zero-dimensional version of the $\sigma $-model for the unitary ensemble is
exactly the same as the zero-dimensional $\sigma $-model derived in Ref.\cite
{fyodorov} for the weakly non-Hermitian matrices. Complex random
non-Hermitian matrices appeared in studies of dissipative quantum maps \cite
{haake,grobe}, which justifies an interest to studying the unitary ensemble.
By the term ``weakly non-Hermitian'' the authors of Ref.\cite{fyodorov}
called matrices $X$ that could be represented in the form \begin{equation} 
X=A+i\alpha N^{-1/2}B \; , \label{a3} \end{equation}
where $A$ and $B$ are Hermitian $N\times N$ matrices and $\alpha $ is a
parameter characterizing the non-Hermiticity.

The $\sigma $-model obtained from the ensemble of the matrices, Eq. (\ref{a3}%
), and from the Hamiltonian with the imaginary vector-potential allows us to
relate the parameters $h$ and $\alpha $ to each other. A similar
correspondence exists for the orthogonal ensemble.

Study of the distributions of the complex eigenvalues revealed a striking
difference between the orthogonal and unitary ensembles. The function $%
P(\epsilon ,y)$, Eq. (\ref{a2}), is a smooth positive function of $y$ for
the unitary (provided the disorder is not very strong, this function does
not depend on $\epsilon $). It reaches its maximum at $y=0$ and monotonously
decays with increasing $y$. The corresponding function $P(\epsilon ,y)$ for
the orthogonal ensemble is a sum of a smooth function and a $\delta $%
-function of $y$. This means that a finite fraction of the eigenvalues
remain real at any degree of the non-Hermiticity.

In all the works done in statistical physics only the orthogonal and unitary
were considered. The symplectic ensembles have not been even mentioned,
apparently due to the absence of any applications. However, for the random
matrix ensembles applied for clarifying properties of QCD models\cite
{ver,janik,stephanov,verbaar}, the symplectic ensemble is of the same
importance as the orthogonal and unitary ones. Moreover, numerical results
for distributions of complex eigenvalues presented in Ref.\cite{verbaar}
demonstrate a pronounced difference between the ensembles. The distribution
of the complex eigenvalues on the complex plane is homogeneous in the case
of the unitary ensemble while it shows an {\it accumulation} of the
eigenvalues along the real axis. This corresponds to the presence of the $%
\delta $-function in the function $P(\epsilon ,y)$, Eq. (\ref{a2}), found in
Ref.\cite{efetov}. Although the authors of Ref.\cite{verbaar} considered
chiral matrices, the dependence of the number of eigenvalues at the real
axis on a parameter characterizing the non-Hermiticity was found to be
exactly the same as in Ref.\cite{efetov}. This shows that the phenomenon of
the accumulation is quite general.

A completely different behavior was found in Ref.\cite{verbaar} for the
symplectic ensemble. The distribution function of the complex eigenvalues is
in this case smooth but the probability of real eigenvalues turns to zero,
which corresponds to a {\it depletion} of the eigenvalues along the real
axis. This is a new effect that clearly motivates an analytical
investigation of non-Hermitian symplectic matrices.

In the present publication the distribution function $P(\epsilon ,y)$, Eq. (%
\ref{a2}), is calculated for the ensemble of symplectic non-Hermitian
matrices. This is done by writing a proper zero-dimensional $\sigma $-model.
We are able to obtain an explicit expression for the function $P(\epsilon
,y) $ and demonstrate the depletion of the eigenvalues along the real axis.

The paper is organized as follows: In Sec. II, we introduce the notations
and remind to the reader the scheme of the derivation of the $\sigma $%
-model. In Sec. III, we present the parametrization of the supermatrices $Q$
for the symplectic ensemble. In Sec. IV, the joint probability density of
complex eigenvalues is calculated. Sec. V summarizes the results, and in the
Appendix the Jacobian of the parametrization is derived.

\section{Non-linear $\sigma $-model}

The derivation of the $\sigma $ model for the non-Hermitian orthogonal and
unitary ensembles has been comprehensively presented in Ref. \cite{efetov}.
Addressing to this paper for all details, we repeat some intermediate steps,
concentrating on minor changes that have to be done in the symplectic case.
The final goal is to derive the joint probability density of complex
eigenenergies $P(\epsilon ,y)$, Eq. (\ref{a2}). Of course, one can derive
the zero-dimensional $\sigma $-model from the ensemble of symplectic random
matrices but we prefer to start from the Hamiltonian, Eq. (\ref{a1}), adding
to it spin-orbit impurities.

Due to the non-Hermiticity the Hamiltonian, the notion of advanced and
retarded Green functions, $G_{\epsilon }^{A}$ and $G_{\epsilon }^{R}$
usually used in perturbation theory and in deriving the non-linear $\sigma $%
-models becomes meaningless since they loose their analytic properties. The
difficulty can be overcome by introducing an Hermitian double size operator $%
\hat{M}$ of the form  \begin{equation} \hat{M}=\left(  \begin{array}{cc}
H^{\prime }-\epsilon & i(H^{\prime \prime }-y) \\ 
-i(H^{\prime \prime }-y) & -(H^{\prime }-\epsilon )
\end{array} \right) , \label{b1} \end{equation} where  \begin{equation}
H^{\prime }=\frac{(H+H^{+})}2 \; ,\ \ \ H^{\prime \prime }=-\frac{i(H-H^{+})}2
\; . \label{b2} \end{equation}

In equations (\ref{b1}) and (\ref{b2}), $H$ is the Hamiltonian, Eq. (\ref{a1}),
and $H^{+}$ means its Hermitian conjugated. Instead of manipulating the
non-Hermitian operator, one can use the Hermitian operator $\hat{M}$\cite
{efetov}. Using the ``effective Hamiltonian'' $\hat{M}$, Eq. (\ref{b1}), one
can represent the complex eigenvalues distribution function $P(\epsilon ,y)$%
, Eq. (\ref{a2}), in a form of a functional integral over supervectors $\psi
\left( {\bf r}\right) $ with the weight $\exp (-{\cal L})$ with the
Lagrangian ${\cal L}$ taking the form  \begin{equation}
{\cal L}=-i\int \overline{\psi }({\bf r})[{\cal H}_{0}+U({\bf r}%
)+V_{so}\left( {\bf r}\right) ]\psi ({\bf r})\,d{\bf r} \;  . 
\label{b3} \end{equation}
Here, $\psi ({\bf r})$ and $\overline{\psi }({\bf r})$ are the standard
supervector and its charge-conjugated counterpart, respectively, composed
from anticommuting and commuting fields \cite{book}. The matrix operator $%
{\cal {H}}_0$ consists of two terms  \begin{equation}
{\cal H}_0=(H_0^{\prime }-\epsilon +i\gamma \Lambda )I+i\Lambda
_1(H_0^{\prime \prime }+y\tau _3) \; , \label{b4} \end{equation}
where $H_0^{\prime }$ and $H_0^{\prime \prime }$ have the form 
\begin{equation}
H_0^{\prime }=\frac{{\bf \hat{p}}^2}{2m} \; ,\qquad H_0^{\prime \prime }=-
i\frac{%
{\bf h\hat{p}}}m \; .  \label{b5} \end{equation}
In equation (\ref{b4}), $\gamma $ is a small positive number that should be put
to zero at the end of calculations. The term $V_{so}$ in Eq. (\ref{b3})
stands for spin-orbit impurities. It can be derived from the initial
Hamiltonian after a formal inclusion of the interaction $U_{so}\left( {\bf r}%
\right) $ with the spin-orbit impurities. The simplest form of this
interaction can be written as follows \begin{equation}
U_{so}({\bf r})=\mbox{\boldmath $\sigma$}\,[\nabla u_{{\rm so}}({\bf r})\times 
\hat{{\bf %
p}}]  \; ,  \label{b6} \end{equation}
where the vector \mbox{\boldmath $\sigma$} is formed from the Dirac matrices $\sigma
_x $, $\sigma _y$, and $\sigma _z$. The matrices $I$, $\Lambda $, $\Lambda _1
$ and $\tau _3$ entering Eq. (\ref{b4}) have the form 
\begin{equation} I=\left(  \begin{array}{cc}
{\bf 1} & 0 \\  0 & {\bf 1} \end{array}
\right) ,\quad \Lambda _1=\left( \begin{array}{cc}
0 & {\bf 1} \\  {\bf 1} & 0 \end{array}
\right)  ,\quad \Lambda =\left(  \begin{array}{cc} {\bf 1} & 0 \\ 
0 & -{\bf 1} \end{array} \right) ,  \label{b7}
\end{equation} \begin{equation} \tau _3=\left( 
\begin{array}{cc} 1 & 0 \\  0 & -1 \end{array}
\right)  .  \label{b8} \end{equation}
Due to the necessity of considering the spin variables the supervectors $%
\psi \left( {\bf r}\right) $ have now $16$ components. The unit blocks ${\bf %
1}$ in the matrices in Eq. (\ref{b7}) have the size $8\times 8$ and unities $%
1 $ entering the matrix $\tau _3$ are $2\times 2$ matrices.

The distribution of the electric fields $\nabla u_{{\rm so}}({\bf r})$ is
assumed to be Gaussian:  \begin{equation}\langle \nabla u_{{\rm so}}({\bf r)}
\rangle =0 \; ,\ \ \ \langle \partial _{i}u_{%
{\rm so}}({\bf r})\,\partial _{j}u_{{\rm so}}({\bf r})\rangle =\frac{\delta
_{ij}\,\delta ({\bf r}-{\bf r}^{\prime })}{6\pi \nu \tau _{{\rm so}}}
\; , \label{b9} \end{equation}
where the density of states at ${\bf h}=0$ at the Fermi surface $\nu
=mp_{0}/2\pi ^{2}$ and $\tau _{{\rm so}}$ is the spin-orbit scattering time.

Further transformations are performed according to standard rules of the
supersymmetry technique\cite{book}. Averaging over the disorder results in
an interaction term $\psi ^4$ in the Lagrangian ${\cal L}$. This term is
decoupled by integration over a supermatrix $Q$. Then, one integrates the
supervector $\psi $ out, arriving thus at an integral over $Q$ with the
weight $\exp (-F[Q])$, where $F[Q]$ is a free energy functional.

The spin-orbit interactions lead to additional ``effective fields'' of a
certain symmetry in the free energy functional $F[Q]$. These fields lower
the symmetry of the functional. As a result, a part of fluctuational modes
have a gap and their contribution at low energies can be neglected. This is
equivalent to putting certain elements of the supermatrix $Q$ to zero.
Carrying out this procedure, one comes to a matrix $Q$ with spin blocks
proportional to unit matrices. This is equivalent to consideration the model
with $8\times 8$ supermatrices $Q$ having a new symmetry. These are the same
supermatrices as those used in Ref.\cite{book} for description of the
symplectic case.

Therefore, we should perform calculations similar to those of Ref.\cite
{efetov} but integrating over the supermatrices $Q$ with the symmetry
corresponding to the symplectic ensemble. After standard transformations, we
reduce the distribution function $P(\epsilon ,y)$ to the following integral 
\begin{equation}
P(\epsilon ,y)=-\frac{\pi \nu }{4\Delta }\int A[Q]\exp \left( -F[Q]\right)
dQ \; ,  \label{b10} \end{equation} \[
A[Q]=\left( Q_{42}^{11}+Q_{42}^{22}\right) \left(
Q_{24}^{11}+Q_{24}^{22}\right) -\left( Q_{42}^{21}+Q_{42}^{12}\right) \left(
Q_{42}^{21}+Q_{24}^{12}\right) \]
with the zero-dimensional version of the free-energy functional
\begin{equation}
F[Q]=STr\left( \frac{a^{2}}{16}[Q,\Lambda _{1}]^{2}-\frac{x}{4}\Lambda
_{1}\tau _{3}Q\right) \; .  \label{b11} \end{equation}
In equation (\ref{b11}), the symbol $[..]$ stands for commutator, $STr$ for
supertrace and we have introduced the following parameters: \begin{equation}
a^{2}=\frac{2\pi D_{0}h^{2}}{\Delta } \; ,\qquad x=\frac{2\pi 
y}{\Delta } \; , \label{b12} \end{equation}
where $D_{0}$ is the classical diffusion coefficient and $\Delta =(2\nu
V)^{-1}$ is the mean level spacing (the factor $2$ in this expression is due
to lifting of the spin degeneracy by the spin-orbit impurities).

\section{Parametrization for the supermatrices $Q$}

To integrate over all symplectic matrices, proper variables parametrizing
the $Q$ supermatrix should be introduced. The parameters have to be chosen
so that to cover all the set of the symplectic matrices: Any symplectic
matrix has to be reached only once.

Although the parametrization for the unitary and the orthogonal ensembles
cannot be used for the symplectic ensemble only minor changes have to be
done to adjust the non-Hermitian parametrization \cite{efetov} to the case
under consideration. As in Ref.\cite{efetov}, we represent the $Q$ matrix
in the form of the product  \begin{equation}
Q=TYQ_0\overline{Y}\,\overline{T} \; .  \label{c1} \end{equation}

To fulfill the constrain $Q^{2}=1$, the following equalities must be
hold \begin{equation} Q_{0}^{2}=1\; ,\ \ T\overline{T}=1\; ,\ \ 
Y\overline{Y}=1 \; . \label{c2} \end{equation}

As in Ref.\cite{efetov}, the supermatrices $T$ and $Y$ are chosen to commute
with $\Lambda _{1}$%
\begin{equation} \lbrack T,\Lambda _{1}]=[Y,\Lambda _{1}]=0 \; . \label{c3}
\end{equation}

The next simplification facilitates the parametrization and enables one to
calculate the Jacobians quickly. Namely, we decompose the supermatrix $Y$
into the product of the matrix $Y_{0}$ containing commuting variables and
the matrices $R$ and $S$, consisting of the Grassmann ones 
\begin{equation} Y=Y_{0}RS \; . \label{c4} \end{equation}

The $2\times 2$ blocks in the matrices $R$, $S$, and $Q_{0}$ are chosen to
be diagonal; the necessary symmetry of the $2\times 2$ blocks $a$, $b$ and $%
\sigma $ is achieved by a proper choice of $2\times 2$ blocks of the matrix $%
Y_{0}$. Thus, the matrices $Q_{0}$, $R$, and $S$ can be written in a form
similar to the one in \cite{efetov} 
\begin{equation} Q_{0}=\left(  \begin{array}{cc}
\cos \hat{\varphi} & -\tau _{3}\sin \hat{\varphi} \\ 
-\tau _{3}\sin \hat{\varphi} & -\cos \hat{\varphi}
\end{array} \right) ,\ \ \hat{\varphi}=\left( 
\begin{array}{cc} \varphi & 0 \\  0 & i\chi \end{array}
\right) , \label{c5} \end{equation}
where $\varphi $ and $\chi $ are proportional to the unity $2\times 2$
matrix,  \begin{equation} R=\left(  \begin{array}{cc} \hat{R} & 0 \\ 
0 & \hat{R} \end{array} \right) ,\ \ \hat{R}=\left( 
\begin{array}{cc} 1-2\rho \overline{\rho } & 2\rho \\ 
-2\overline{\rho } & 1+2\rho \overline{\rho }
\end{array} \right) , \label{c6} \end{equation} and 
\begin{equation} S=\left( 
\begin{array}{cc} 1-2\hat{\sigma}^{2} & 2i\hat{\sigma} \\ 
2i\hat{\sigma} & 1-2\hat{\sigma}^{2}
\end{array} \right) ,\ \ \hat{\sigma}=\left( 
\begin{array}{cc} 0 & \sigma \\ 
\overline{\sigma } & 0 \end{array}
\right) .  \label{c7} \end{equation}

The matrices $\rho $ and $\sigma $ in Eqs. (\ref{c6}, \ref{c7}) have the
form  \begin{equation} \rho =\left(  \begin{array}{cc} \rho & 0 \\ 
0 & \rho ^{*} \end{array} \right) ,\ \ \sigma =\left( 
\begin{array}{cc} \sigma & 0 \\  0 & \sigma ^{*}
\end{array} \right).  \label{c8} \end{equation}

The next step is to represent the supermatrix $Y_{0}$ in Eq. (\ref{c4}) as the
product  \begin{equation} Y_{0}=Y_{3}Y_{2}Y_{1} \; ,  \label{c9}
\end{equation} where $Y_{3}$ is the diagonal matrix  \begin{equation}
Y_{3}=\left( \begin{array}{cc} \exp (i\hat{\beta}/2) & 0 \\  0 & 
\exp (i\hat{\beta}/2) \end{array} \right) ,\ \ \hat{\beta}=\left( 
\begin{array}{cc} \beta \tau _{3} & 0 \\  0 & \beta _{1}\tau _{3}
\end{array} \right) . \label{c10} \end{equation}

In order to recover the symplectic symmetry we have to choose the matrices $%
Y_{1}$ and $Y_{2}$ as follows  \begin{eqnarray} Y_{1} &=&\left( 
\begin{array}{cc} \hat{w} & 0 \\  0 & \hat{w} \end{array}
\right) ,\ \ \hat{w}=\left(  \begin{array}{cc} 1 & 0 \\ 
0 & w \end{array} \right) ,  \label{c11} \\ w &=&\left( 
\begin{array}{cc} \cosh (\mu /2) & -i\sinh (\mu /2) \\ 
i\sinh (\mu /2) & \cosh (\mu /2)
\end{array} \right) ,  \nonumber \end{eqnarray}
and  \begin{eqnarray} Y_{2} &=&\left( 
\begin{array}{cc}
\cos (\hat{\theta}_{2}/2) & -i\sin (\hat{\theta}_{2}/2) \\ 
-i\sin (\hat{\theta}_{2}/2) & \cos (\hat{\theta}_{2}/2)
\end{array} \right) ,  \label{c110} \\
\hat{\theta}_{2} &=&\left( 
\begin{array}{cc} \theta _{2}\tau _{1} & 0 \\  0 & 0
\end{array} \right) ,\ \ \tau _{1}=\left( 
\begin{array}{cc} 0 & 1 \\  1 & 0 \end{array}
\right) . \nonumber \end{eqnarray}
Finally, the supermatrix $T$ can be taken as 
\begin{equation} T=\left(  \begin{array}{cc}
u & 0 \\  0 & u \end{array} \right) \left( 
\begin{array}{cc}
\cos (\hat{\theta}/2) & -i\sin (\hat{\theta}/2) \\ 
-i\sin (\hat{\theta}/2) & \cos (\hat{\theta}/2)
\end{array} \right) \left( 
\begin{array}{cc} v & 0 \\  0 & v \end{array} \right) ,  \label{c12}
\end{equation}
where  \[ \hat{\theta}=\left(  \begin{array}{cc} \theta & 0 \\ 
0 & i\theta _{1} \end{array} \right) ,  \]
\begin{equation} u=\left(  \begin{array}{cc}
1-2\eta \overline{\eta } & 2\eta \\ 
-2\overline{\eta } & 1-2\overline{\eta }\eta
\end{array} \right) ,\ \ v=\left( 
\begin{array}{cc}
1-2\kappa \overline{\kappa } & 2\kappa \\ 
-2\overline{\kappa } & 1-2\overline{\kappa }\kappa
\end{array} \right) . \label{c120} \end{equation}

The $2\times 2$ matrices $\theta $ and $\theta _{1}$ in Eq. (\ref{c12}) are
proportional to the unit matrix and the matrices $\eta $ and $\kappa $ are 
\[ \eta =\left( \begin{array}{cc} \eta & 0 \\  0 & \eta ^{*}
\end{array} \right) ,\ \ \kappa =\left( 
\begin{array}{cc} \kappa & 0 \\  0 & \kappa ^{*}
\end{array} \right) . \]
The explicit form of the supermatrix $Q$ within the parametrization
suggested, Eqs. (\ref{c5}-\ref{c120}), is very similar to that for the
orthogonal ensemble\cite{efetov} and differs from the latter by minor
changes in the matrices $\hat{w}$, $\hat{\theta}_{2}$, $\sigma $, $\rho $, $%
\eta $, and $\kappa $.

To ensure the unambiguity of the parametrization, we should specify the
variation range of the variables. This is done by comparing the compact and
the noncompact sector with those in the standard parametrization \cite{book}%
. As the result, the variables vary in the following intervals: 
\begin{eqnarray}
-\pi /2 &<&\varphi <\pi /2\; ,\ \ 0<\chi <\infty \; ,\ \ -\pi <\theta <\pi \;,\ \
-\infty <\theta _{1}<\infty \;,  \label{c13} \\
0 &<&\mu <\pi \; ,\ \ 0<\theta _{2}<\infty \; ,\ \ 0<\beta <\pi \; ,\ \ 0<\beta
_{1}<2\pi \; . \nonumber \end{eqnarray}

The only thing that remains to be done to perform explicit calculations for
physical quantities is calculation of the Jacobian of the transformation to
the variables described by Eqs. (\ref{c5}-\ref{c120}).

Its derivation presented in the Appendix leads to the following final result
for the elementary volume  \begin{equation}
\lbrack dQ]=J_\varphi J_\theta J_\mu J_cdR_B\,dR_F \; ,  \label{c130}
\end{equation} where  \begin{eqnarray}
J_\varphi &=&\frac 1{8\pi }\frac{\cos \varphi \cosh \chi }{(\sinh \chi
+i\sin \varphi )^2}\; ,\ \ J_\theta =\frac 1{32\pi }\frac 1{\sinh ^2\frac 12%
(\theta _1+i\theta )} \; ,  \label{c14} \\
J_\mu &=&\frac 1{2^8\pi ^2}\frac{\sin \theta _2\sinh \mu }{(\cos \theta
_2-\cosh \mu )^2}\; ,\ \ J_c=\frac{4\sinh ^2\chi }{(\sinh \chi -i\sin \varphi
)^2} \; . \nonumber \end{eqnarray}
and 
\begin{equation}
dR_B=d\theta \,d\theta _1d\varphi \,d\chi \,d\mu \,d\theta _2d\beta \,d\beta
_1\; ,\ \ dR_F=d\eta \,d\eta ^{*}d\kappa \,d\kappa ^{*}d\sigma \,d\sigma
^{*}d\rho \,d\rho ^{*} \; . \label{c140}
\end{equation}

Equations (\ref{c1}-\ref{c14}) are sufficient for evaluation of any integral over
the supermatrices $Q$ and, with the help of Eqs. (\ref{b10}, \ref{b11}),
provides a straightforward way of calculating the distribution function of
complex eigenvalues $P(\epsilon ,y)$, Eq. (\ref{a2}).

\section{Density of complex eigenvalues}

Before staring the calculations let us introduce more compact notations. As
it will be seen in what follows, only the following combinations of the
variables describing the parametrization of the supermatrix $Q$ enter all
functions of interest  \begin{equation}
t=\sin \varphi \; ,\ \ z=\sinh \chi \; ,\ \ \omega =\cosh \mu \; ,\ \ \lambda 
=\cos \theta _{2} \; .  \label{d1} \end{equation}

Since the matrices $T$ and $Y$ commute with $\Lambda _{1}$, the first term
in the free energy, Eq. (\ref{b11}) does not depend on them. The second term
in Eq. (\ref{b11}) does not depend on $T$. As a result, the free energy $%
F[Q] $ takes a rather simple form  \begin{equation}
F[Q]=a^{2}(t^{2}+z^{2})+x[(\lambda t-i\omega z)+4(\sigma \sigma ^{*}+\rho
\rho ^{*})(\omega -\lambda )(t-iz)] \; . \label{d2} \end{equation}

The fact that $F[Q]$ does not depend on $T$ simplifies the integration over $%
Q$ in Eq. (\ref{b10}). Using the parametrization, Eqs. (\ref{c1}\ref{c12}),
we can also represent the supermatrix $Q$ as  \begin{equation}
Q=u\widetilde{Q}\overline{u}  \; , \label{d3} \end{equation}
with $u$ from Eq. (\ref{c120}) and some supermatrix $\widetilde{Q}$.
Substituting Eq. (\ref{d3}) into Eq. (\ref{b10}) for the density of complex
eigenvalues $P(\epsilon ,y)$ and integrating over $\eta $ and $\eta ^{*}$,
one represents this function in the form  \begin{eqnarray} P(\epsilon ,y) 
&=&\frac{\pi \nu }{4\Delta }\int [Str(\tau _3\Lambda _1%
\widetilde{Q})]^2\,\exp (-F[Q])\,d\widetilde{Q}  \label{d4} \\
&=&\frac{4\pi \nu }\Delta \frac{d^2}{dx^2}\int \exp (-F[Q])\,dQ \; . 
\nonumber \end{eqnarray}

For the symplectic ensemble, one has in Eq. (\ref{d4}) an uncertainty of the
type $0\!\times \!\infty $, since the integrand does not contain the
variables $\kappa $ and $\kappa ^{*}$ and, on the other hand, the Jacobians $%
J_{\theta }$ and $J_{\mu }$ are singular as $\theta $, $\theta _{1}$, $%
\theta _{2}$, and $\mu \rightarrow 0$. To resolve this singularity, we can
use the regularization procedure developed for the orthogonal ensemble \cite
{efetov}. All the manipulations are identical to those of Ref.\cite{efetov},
because the free energy, Eq. (\ref{d2}) has the form similar to that of the
orthogonal ensemble. Moreover, the singularities of the Jacobians, Eqs. (\ref
{c14}), are the same as the ones for the orthogonal ensemble. We do not
specify the procedure once more and present here only the final result of
the regularization with proper changes of notations. The function $%
P(\epsilon ,y)$ can be written in the form of a sum of two terms 
\begin{equation} P(\epsilon ,y)=P^{(1)}(\epsilon ,y)+P^{(2)}(\epsilon ,y) \; ,  
\label{d5} \end{equation} where 
\begin{equation}
P^{(1)}(\epsilon ,y)=\frac{\nu }{4\Delta }\frac{d^{2}}{dx^{2}}\int \exp
[-a^{2}(t^{2}+z^{2})-x(t-iz)]\,\frac{4z^{2}dtdz}{(t^{2}+z^{2})^{2}}
\label{d6} \end{equation} 
and  \begin{eqnarray}
P^{(2)}(\epsilon ,y) &=&\frac{\nu }{4\Delta }\frac{d^{2}}{dx^{2}}\int \exp
[-a^{2}(t^{2}+z^{2})-x(t\omega -i\lambda z)]  \label{d7} \\
&&\times \frac{(t-iz)^{2}\,z^{2}x^{2}}{(t^{2}+z^{2})^{2}}\,dt\,dz\,d\omega
\,d\lambda  \; . \nonumber \end{eqnarray}
The integration in Eqs. (\ref{d6}) and (\ref{d7}) is performed in the
intervals $-1<t<1$, $-\infty <z<\infty $, $1<\omega <\infty $, and $%
-1<\lambda <1$.

To perform the integration in Eq. (\ref{d7}) over $\lambda $, one should
introduce an infinitesimal positive $\delta $, defining $z_{-}$ according to 
$z_{-}=z+i\delta \,{\rm sgn}(x)$, so that the integral becomes convergent.
After this, the integration over $\lambda $ and $\omega $ in Eq. (\ref{d7})
is easily carried out. Adding the result of the integration to Eq. (\ref{d6}%
) we obtain  \begin{eqnarray}
P(\epsilon ,y) &=&\frac{\nu }{4\Delta }\,\frac{d^{2}}{dx^{2}}%
\int_{-1}^{+1}dt\,\int_{-\infty }^{+\infty }dz\exp
(-a^{2}(t^{2}+z_{-}^{2}))[(t+iz_{-})^{2}\exp (ixz_{-}-tx)  \label{d8} \\
&&-(iz_{-}-t)^{2}\exp (ixz_{-}+tx)]\,\frac{z_{-}}{it(t^{2}+z_{-}^{2})^{2}} \; .
\nonumber \end{eqnarray}

Comparing Eq. (\ref{d8}) with its analog for the orthogonal ensemble, we
notice an important difference between them: The variable $z_{-}$ is present
in the numerator in Eq. (\ref{d8}), whereas it stands in the denominator of
the equation for the orthogonal ensemble. In the latter case, the
distribution function $P(\epsilon ,y)$ contains an additional contribution
of a $\delta $-function after differentiation of $z_{-}$ over $x$. For the
symplectic ensemble, the differentiation of $z_{-}$ in the numerator leads
to no singularity on the real axis and one take the limit $\delta
\rightarrow 0$ before calculating the integral in Eq. (\ref{d8}). Thus, only
the exponents should be differentiated over $x$. These differentiations
simplifies considerably the integrand and the integration over $z$ can be easily
carried out. After that one obtains  \begin{equation}
P(\epsilon ,y)=\frac \nu \Delta \,\frac x{4a^3}\,\sqrt{\pi }\,\exp \left( -%
\frac{x^2}{4a^2}\right) \int_0^1dt\exp (-a^2t^2)\,\frac{\sinh (tx)}t \; .
\label{d9} \end{equation}

Equation (\ref{d9}) solves completely the problem involved and is the main 
result of the present paper.

The following properties of the density function $P(\epsilon ,y)$ are easily
checked: It is symmetric with respect to $y$ and is properly normalized 
\begin{equation} \int dy\,P(\epsilon ,y)=1 \; . \label{d10} \end{equation}

In the limit $a\gg 1$(the limit of a strong non-Hermiticity), one obtains
the universal asymptotics valid for all three ensembles 
\begin{equation}
P(\epsilon ,y)\simeq \frac{\pi \nu }{2a^2\Delta }\left\{ 
\begin{array}{cc} 1, & 2a\ll |x|<2a^2 \\ 
0, & |x|>2a^2 \end{array} \right.  \label{d11} \end{equation}

The form of the density of complex eigenstates, Eq. (\ref{d11}), corresponds%
\cite{efetov} to the ``elliptic law'' of Refs.\cite{ginibre,girko}.

In the opposite limit $a\ll 1$, the function $P(\epsilon ,y)$ can be written
as  \begin{equation}
P(\epsilon ,y)=\frac \nu \Delta \frac{x^2}{4a^3}\sqrt{\pi }\exp \left( -%
\frac{x^2}{4a^2}\right) \; . \label{d12}
\end{equation}

The behavior of the function $P(\epsilon ,y)$, Eq. (\ref{d9}), at small $y$
(related to $x$ by Eq. (\ref{b12})) is drastically different from the
behavior of the corresponding functions for the orthogonal and unitary
ensembles\cite{efetov}. This function is small at small $y$, being
proportional to $y^2$ and turns to zero in the limit $y\rightarrow 0$. This
means that the probability that eigenvalues remain real at finite degree of
the non-Hermiticity is zero. In other words, the distribution function of
complex eigenvalues exhibits a depletion along the real axis. The depletion
region broadens with increasing the non-Hermiticity. The function $%
P(\epsilon ,y)$ is represented in Fig. 1 for several values of $a\approx 1$.

\section{Conclusions}

In the present paper, we studied analytically disordered non-Hermitian
models with the symplectic symmetry. This is the last of three universality
classes that has not been considered yet. Using the supersymmetry technique
we derived a proper non-linear $\sigma $-model starting from the a model of
disorder with a direction. The zero-dimensional version of the non-linear $%
\sigma $-model corresponds the ensemble of random non-Hermitian symplectic
matrices. Within the zero-dimensional $\sigma $-model, we calculated the
joint probability density function of complex eigenvalues. We introduced a
convenient parametrization and calculated the Jacobian corresponding to this
parametrization.

All this allowed us to derive an explicit expression for the density of
complex eigenvalues. Asymptotic behavior of this function demonstrates
clearly that the basic properties of the system depend strongly on the
ensemble. Introducing the non-Hermiticity in the Hamiltonian affects very
differently the spectrum of three ensembles. Only when the non-Hermiticity
is very large, the difference is no longer important.

It is known from previous works that the eigenvalues of a system belonging
to the unitary ensemble are smoothly distributed around the real axis. The
density function for the orthogonal ensemble contains a $\delta $-function
contribution on the real axis describing an accumulation of the eigenvalues.
In contrast to the previous cases, we obtained for the symplectic ensemble a
depletion of eigenvalues along the real axis, which is in a good agreement
with the results of a numerical study \cite{verbaar}. These features
correspond to a tendency of a system from the orthogonal ensemble to
preserve localized behavior. However, after introducing spin-orbit
impurities, the system acquires delocalized features.

\section{Appendix}

The Jacobian of the parametrization specified by Eqs. (\ref{c1}-\ref{c120})
can be derived from the elementary length $Str(dQ)^{2}$. The most economical
way to proceed is to compare the parametrization involved with that for the
orthogonal ensemble \cite{efetov}. Two essential differences are easily
noticed: the $2\times 2$ blocks in matrices $Y_{1}$ and $Y_{2}$ are
interchanged and all the conjugated Grassmann variables have the opposite
sign. The last difference, however, does not lead to any change in the
calculation, so long as the contribution to the length $Str(dQ)^{2}$ from
the Grassmann variables is due to terms of the kind $\overline{\eta }%
\,d\kappa $, $\overline{\eta }\,d\eta $ etc. Taking this into account, we
can immediately reduce the elementary length to the following expression 
\begin{eqnarray}
Str(dQ)^{2} &=&Str((dQ_{0})^{2}+[\delta Z,Q_{0}]^{2})\; ,  \label{f1} \\
\delta Z &=&\overline{S}\,\overline{R}(\overline{Y}_{0}\delta TY_{0}+dR\,%
\overline{R}+R\,dS\,\overline{S}\,\overline{R}+\delta Y_{0})R\,S \; , 
\nonumber \end{eqnarray}
where all the terms apart of the last one, $\overline{S}\,\overline{R}\delta
Y_{0}\,R\,S$, are identical to those in Ref.\cite{efetov}.

Using Eq. (\ref{c9}), we write $\delta Y_{0}$ as 
\begin{equation}
\delta Y_{0}=\delta Y_{1}+\delta Y_{2}+\overline{Y_{1}}\overline{Y_{2}}%
\delta Y_{3}\,Y_{2}\,Y_{1} \; ,  \label{f100} \end{equation}
which can be rewritten in the form 
\begin{eqnarray} \delta Y_{0} &=&{\bf 1}\frac{i}{2}\,\left[ \left( 
\begin{array}{cc} d\beta _{1}\tau _{3}\cos \theta _{2} & 0 \\ 
0 & d\overline{w}\tau _{3}w \end{array} \right) -d\mu \left( 
\begin{array}{cc} 0 & 0 \\  0 & \tau _{2} \end{array}
\right) \right]  \label{f2} \\
&&+\Lambda _{1}\frac{1}{2}\,\left[ -\left( 
\begin{array}{cc} \tau _{2}d\beta _{2}\sin \theta _{2} & 0 \\  0 & 0
\end{array} \right) +d\theta _{2}\left( 
\begin{array}{cc} \tau _{1} & 0 \\  0 & 0 \end{array}
\right) \right] ,  \nonumber \end{eqnarray}
where  \begin{equation} \tau _{2}=\left(  \begin{array}{cc}
0 & -i \\  i & 0 \end{array} \right) . \label{f3} \end{equation}

Multiplying three matrices with each other, one obtains 
\begin{eqnarray}
\overline{Y}_0\delta TY_0 &=&{\bf 1\times }2\left[ \cos \frac{\theta _2}2%
\left(  \begin{array}{cc} 0 & d\kappa ^{\prime } \\ 
-d\overline{\kappa }^{\prime } & 0 \end{array}
\right) +i\sin \frac{\theta _2}2\left( 
\begin{array}{cc} 0 & \tau _1d\eta ^{\prime } \\ 
-d\overline{\eta }^{\prime }\tau _1 & 0
\end{array} \right) \right]  \label{f4} \\
&&+2i\Lambda _1\left[ \cos \frac{\theta _2}2\left( 
\begin{array}{cc} 0 & \eta ^{\prime } \\ 
\overline{\eta }^{\prime } & 0
\end{array} \right) -i\sin \frac{\theta _2}2\left( 
\begin{array}{cc}
0 & \tau _1d\kappa ^{\prime } \\ 
d\overline{\kappa }^{\prime }\tau _1 & 0 \end{array}
\right) -\frac i2d\hat{\theta}\right] , \nonumber \end{eqnarray}
where $d\kappa ^{\prime }=d\kappa w\,\exp [i(\beta -\beta _1)/2]$ and $d%
\overline{\kappa }^{\prime }=\overline{w}d\overline{\kappa }\,\exp [i(\beta
-\beta _1)/2]$ and analogous for $d\eta ^{\prime }$ and $d\overline{\eta }%
^{\prime }$. One should keep in mind that the differentials $d\eta $ and $%
d\kappa $ in Eq. (\ref{f4}) are not the initial variables entering Eq. (\ref
{c120}) but new variables obtained from the initial ones by several
replacements and shifts common for all three ensembles. The Jacobian of
those transformations $J_{\theta \text{ }}$ is given by Eqs. (\ref{c14}).

After that we pick up the differentials of the Grassmann variables
(proportional to the unit matrix), make a shift of the differentials
analogous to the one in Ref.\cite{efetov} and introduce the matrix
differentials  \begin{equation} d\sigma =\left( 
\begin{array}{cc} d\sigma _1 & d\sigma _2 \\ 
-d\sigma _2^{*} & d\sigma _1^{*}
\end{array} \right) ,\ \ d\rho =\left( 
\begin{array}{cc} d\rho _1 & d\rho _2 \\ 
-d\rho _2^{*} & d\rho _1^{*}
\end{array} \right) , \label{f5} \end{equation}
where  \begin{equation} \begin{array}{c}
d\sigma _2=-i\cos \frac{\theta _2}2\sinh \frac \mu 2d\eta -\sin \frac{\theta
_2}2\cosh \frac \mu 2d\kappa ^{*} \\ 
d\sigma _2^{*}=-i\cos \frac{\theta _2}2\sinh \frac \mu 2d\eta ^{*}+\sin 
\frac{\theta _2}2\cosh \frac \mu 2d\kappa \\ 
d\rho _2=-i\cos \frac{\theta _2}2\sinh \frac \mu 2d\kappa +\sin \frac{\theta
_2}2\cosh \frac \mu 2d\eta ^{*} \\ 
d\rho _2^{*}=-i\cos \frac{\theta _2}2\sinh \frac \mu 2d\kappa ^{*}-\sin 
\frac{\theta _2}2\cosh \frac \mu 2d\eta \end{array} \label{f6} \end{equation}
The Jacobian of the transformation, Eq. (\ref{f6}), from $\eta $, $\kappa $,
to $\sigma $ and $\rho $, equals  \begin{equation}
\widetilde{J}_\mu =\frac 4{(\cos \theta _2-\cosh \mu )^2}\;  . \label{f60}
\end{equation}

The supermatrix $\delta Z$ from Eq. (\ref{f1}) can be represented as 
\begin{equation} \delta Z=\delta Y_0^{\prime }+i\Lambda _1(2d\hat{\sigma}-
d\hat{\theta}/2)+%
{\bf 1\times }\,2k\,d\hat{\rho} \; ,  \label{f10} \end{equation}
where  \[ k=\left( 
\begin{array}{cc} 1 & 0 \\  0 & -1 \end{array} \right)  \]
and  \begin{eqnarray}
\delta Y_0^{\prime } &=&-{\bf 1}\frac i2\left[ d\beta \,\sinh \mu \left( 
\begin{array}{cc} 0&0 \\ 0&\tau _1 \end{array} \right) \right] +d\mu \left( 
\begin{array}{cc} 0 & 0 \\  0 & \tau _2 \end{array} \right)  \label{f8} \\
&&+\Lambda _1\frac 12\left[ -d\beta _1\,\sin \theta _2\left( 
\begin{array}{cc} \tau _2&0 \\ 0&0 \end{array}\right) +d\theta _2\left( 
\begin{array}{cc} \tau _1 & 0 \\  0 & 0 \end{array}
\right) \right] .  \nonumber \end{eqnarray}
Calculating the length $Str(dQ)^2$ one notices that the anticommuting part
of $\delta Z$, proportional to $\Lambda _1$, decouples from the commuting
one, proportional to the unit matrix ${\bf 1}$. The contribution to the
length from the first part is the same as that of Ref. \cite{efetov} leading
to the Jacobian  \begin{equation}
J_{\varphi \chi }=\frac 1{2^{24}}\frac 1{(\sin ^2\varphi +\sinh ^2\chi )^2} \; ,
\label{f9} \end{equation}
whereas the second part of the elementary length equals  \begin{eqnarray}
Str[\delta Z_{\parallel },Q_0]^2 &=&4\{[(d\mu )^2+(d\beta )^2\sinh \mu
]\sinh ^2\chi +(d\theta )^2\cos ^2\varphi  \label{f20} \\ &&+(d\theta _1)^2\cosh 
^2\chi +(d\theta _2)^2+(d\beta _1)^2\sin ^2\theta _2\} \; . \nonumber
\end{eqnarray}
Since in our parametrization the blocks from the commuting variables in the
matrices $Q_0$ and $T$ are the same as in Ref.\cite{efetov}, the Jacobian $%
J_\theta $ does not change. Combining the contributions from Eqs. (\ref{f60}%
, \ref{f9}, \ref{f20}) with $J_\theta $, we arrive at the elementary volume,
Eq. (\ref{c130}--\ref{c140}).

\newpage 
\centerline{\large  FIGURE CAPTIONS}
\bigskip

Figure 1.
The dependence of complex eigenvalues of energy $P(\epsilon, y)$ on the 
imaginary part of energy $y=x\Delta/2\pi$, for several values of
non-Hermiticity $a=2\pi D_0 h^2/\Delta$.                                   


\begin{center}
  \unitlength1cm
  \begin{minipage}[t]{12cm}
  \begin{picture}(0,5)
     \put(-4.,-24.5)
    {\includegraphics{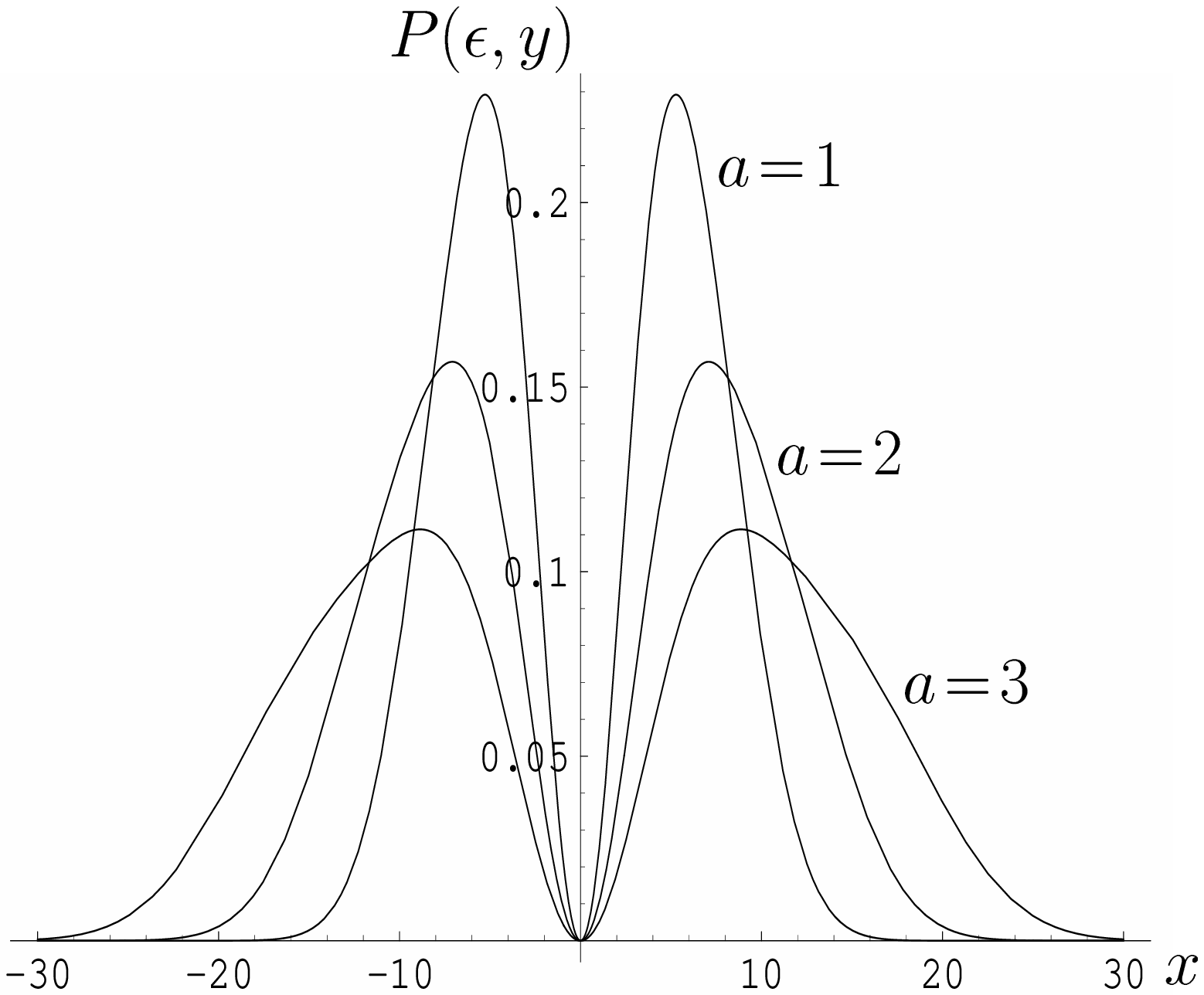}}
 \end{picture}
  \end{minipage}
\end{center}



\end{document}